# A Gait Triaging Toolkit for Overlapping Acoustic Events in Indoor Environments

Summoogum. Kelvin, *Senior Member, IEEE, MiiHealth, UK*
Das. Debayan, *MiiCare, UK*
Jayakumar. Parvati, *MiiCare, UK*

*Abstract* — Gait has been used in clinical and healthcare applications to assess the physical and cognitive health of older adults. Acoustic based gait detection is a promising approach to collect gait data of older adults passively and non-intrusively. However, there has been limited work in developing acoustic based gait detectors that can operate in noisy polyphonic acoustic scenes of homes and carehomes. We attribute this to the lack of good quality gait datasets from the real-world to train a gait detector on. In this paper, we put forward a novel machine-learning based filter which can triage gait audio samples suitable for training machine learning models for gait detection. The filter achieves this by eliminating noisy samples at an $f(1)$ score of 0.85 and prioritising gait samples with distinct spectral features and minimal noise. To demonstrate the effectiveness of the filter, we train and evaluate a deep learning model on gait datasets collected from older adults with and without applying the filter. The model registers an increase of 25 points in its $f(1)$ score on unseen real-word gait data when trained with the filtered gait samples. The proposed filter will help automate the task of manual annotation of gait samples for training acoustic based gait detection models for older adults in indoor environments.

## I. INTRODUCTION

Gait has been extensively studied as a clinical biomarker for assessing deterioration of physical mobility [1] and cognitive health in older adults [2]. Detecting gait patterns of older adults has been explored using accelerometric sensors [3], pressure-sensitive floors [4], video feed [5], even audio recordings [6]. Acoustic based gait detection systems have the added benefit of being non-intrusive, portable and economical when compared to those proposed in contemporary literature for gait analysis. However, there has been little to no progress in this field in the past decade. The advent of large-scale deep learning based techniques have led to the development of state-of-art gait analysis using all of the above data collection modalities except for that using audio.

Acoustic Gait Detection is an example of an Audio Event Detection (AED) task. Our analysis of the literature available for Audio Event Detection (AED) reveal the following findings:

1. Deep learning excels in achieving state of art [7] in AED only when the audio dataset used is clean and contains little to no background noise.
2. Majority of the available works in AED, including those for AGA use data captured from normal healthy individuals in controlled laboratories with clean acoustic scenes [8]. This is not representative of the real-world acoustic scene in an indoor setting with an older adult living with one or more comorbidities. Real world gait recordings are noisy and have multiple overlapping acoustic events (i.e polyphony).
3. There is a very limited number of publicly available annotated acoustic datasets of gait data large enough to train deep learning models for gait detection in the real world.

Manual annotation of real-world gait data helps improve deep learning and machine learning based gait detection models in this regard. However, using manual labour to annotate gait data is time-consuming, not to mention the auditory expertise required to differentiate distinct "good" gait samples from noisy "bad" gait samples in noisy recordings with overlapping acoustic events. Figure 1 below shows the melspectrograms of two samples which a human annotator confidently labelled as "gait" samples despite the dominating presence of background noise and other acoustic events in one of them. Using distinct gait samples during training models will help them learn features better which will automatically translate to more robust detection of gait from noisy polyphonic recordings.

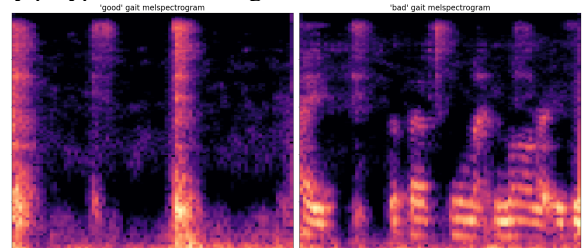

*Figure 1 The two versions of gait as captured in real life*

In this paper, we explore the feasibility of using an automated "filter" which eliminates "bad" gait samples from a population of human-annotated gait samples, leaving us with samples which are most likely to be "good". We take this specific approach because it is much easier to find noisy samples than clean samples from real world recordings. Our objective is to develop a technique that can be deployed in acoustic gait recording devices within the indoor home environment of older adults that will eliminate collection of "bad" gait samples which is unsuitable for training gait detection models, as much as possible directly at source.

This paper summarises the approach we took to develop this gait filter from an pre-annotated gait dataset of footsteps from older adults living with dementia (PLWD) with medium to high fall risks across carehomes in the UK. We present the use of a machine learning model as a "audio signal filter" wherein we train this model on subset of gait samples which we consider as distinct "good" gait samples, use it to classify a test set of gait samples, eliminate the "bad" samples and then evaluate its "filtering" performance against human experts in gait analysis.

In Section II, we discuss the techniques presented in relevant acoustic based gait analysis literature for cleaning and feature extraction from gait samples. In Section III, we explain our process of extracting features to distinguish "good" gait samples from the "bad" noisy ones and passing these features into an ensemble model found by AutoML [9] which maximises detection rate for the "bad" gait samples. In Section IV, we reuse our previous work [6] on a recurrent deep learning model for acoustic based gait detection for older adults and demonstrate how applying the ensemble model as an audio filter improved the detection performance of the deep learning model on real world gait data.

## II. BACKGROUND

### A. Feature Extraction for Gait

It is important to note the distinction between gait and footsteps. Gait is defined as the characteristics in the walking pattern of an individual. Gait "datasets" record sequence of multiple individual footstep events one after the other within a short time window. The majority of existing literature AGA focuses on examination of the individual footstep audio event first and then applying it on all such events in a gait sample. However, there is no available literature (yet) to our knowledge that attempts differentiating clean from noisy gait sounds. The literature instead focuses on characteristics of the audio ("features") which maximise the difference between gait and other non-gait sound like speech. We group these into two camps broadly speaking.

The first camp focuses on using statistical analysis of footsteps in the temporal and frequency domains. This includes extraction of features [10] like power, amplitude, ZCR, peaks, etc. from the waveform and FFT representations. Before the emergence of neural network based feature extraction, such techniques enjoyed popularity across several works in machine learning and signal processing literature for gait detection and analysis. The common shortcoming across all these works using these type of features is that they use gait datasets collected in controlled acoustic scenes [8, 10] with little to minimal background noise. These features are highly susceptible to presence of noise and hence cannot be relied upon in a discusssion on gait detection in the real world with noisy environments.

The second camp relies on spectrogram-based features to distinguish between footstep and other acoustic events. MFCC and melspectrogram by far is the most popular spectrogram feature for deep learning models to learn from. Deep learning models exclusively use neural networks paired with preprocessing layers like a convolution layer (Convolutional Neural Networks) or a reccurent gate operation (Long Short Term Memory and Gated Recurrent Units for instance) to automatically learn the best features to learn from the spectrogram representations. Contemporary AED have achieved state of art [7] by applying deep learning models on spectrograms. This is the most promising among all approaches to build new feature extraction techniques for gait detection, as of the time of writing of this paper.

### B. State of the Art in Gait Detection

Before the advent of deep neural networkks, AGA relied on classifiers like Support Vector Machines (SVM) [10], Gaussian Mixture Models (GMM) [11], Hidden Markov Models (HMM) [12], Non-Negative Matrix Factorizations (NMF) [13] and so on. Annual competitions like DCASE [14] helped popularise the use of use of large deep learning models to achieve state of art in AED tasks like Acoustic Scene Classification, General AED in indoor environments, Audio Tagging, etc. In our previous work [6], we presented the use of synthetic soundscapes to combat the problem of unavailability of gait datasets and trained an Attention-based [15] Bidirectional LSTM model to perform gait detection in noisy overlapping audio streams. This approach was able to reproduce state of art achieved in DCASE 2021 with publicly available gait datasets like TUM GAID [16] and ESC50 [17] in the specific task of gait detection for older adults in indoor environments. However, we found that it failed to detect footsteps when applied on real life gait recordings. In the upcoming section, we discuss what we observed and how we developed the filter as a way to make the model work again.

## III. METHODOLOGY

Our objective in this section is to present the problem we found when we attempted applying our gait detector [6] on a real life gait dataset collected from older adults living in home and carehome settings and our solution to making it work by using a machine learning based audio signal filter to remove gait samples unfit for training the detector.

### A. Dataset Collection

Our primary audio recording device for data collection from the homes of older adults was the MiiCube manufactured by MiiCare [18]. It is a ARM4 processor coupled with a 2-microphone setup from Seed Studio's ReSpeaker Pi Hat [19] enclosed in a 3D-printed cubical box enclosure as shown in Figure 2 below. The microphones are located immediately beneath the top lid to ensure it is capturing the acoustic scene and all acoustic events without exposing the computing unit.

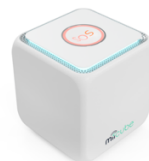

*Figure 2 The MiiCube recording device from MiiCare*

Each MiiCube samples environmental sound at 16 KHz. We chose 16 KHz as the sampling rate because we have observed that gait do not have significant spectral presence beyond 8 KHz (Nyquist frequency). It records audio in batches of 30s at any point of time, and uploads it to a secure cloud storage. We deployed a total of 10 MiiCubes across different carehomes and home sites in the UK. 8 of them were set up in the rooms of older adults in carehomes living with advanced dementia and high fall risk. The remaining 2 were setup in the homes of two different older adults aged 80+ with mild cognitive impairment (MCI). We introduced an additional step of acoustic thresholding based on RMS of the incoming audio streams to determine when to start recording gait. This thresholding was done empirically by manual experimentation, so each MiiCube has its RMS threshold specific to its acoustic scene at the carehome or home where it is deployed.

*B. Human Annotation*

We employed external annotation services to annotate the collected recordings in cloud storage. For this work, we were able to get 6852 audio recordings annotated across 2 acoustic classes: "gait" and "non-gait". For each file, the annotators produced sets of time segments where gait is heard through auditory inspection. The remaining segments of the file were marked as "non-gait". These segments consisted of background noise and from conversations, electrical appliances, door shuts, etc. Both the gait and non-gait segments had overlapping presence of multiple acoustic events. Majority of the gait segments had overlapping background noise and speech. Such quality of sound is to be expected for a real life audio dataset collected from the indoors of homes where individuals lived unlike those in literature [8, 10] collected in controlled laboratory environments.

*C. Motivation behind the filter*

We reproduced our previous work [6] on this newly annotated gait dataset. The preprocessing algorithm generated 3509 melspectrograms in total with 1399 "gait" and 2110 "non-gait" melspectrograms. We applied a 80-20 split to create a training and test set which had 1119 and 280 gait melspectrograms respectively. The model achieved a $f(1)$ score of 0.58 on the 280 gait melspectrograms in the test set. We went through the test set ourselves to see how accurately we as humans could classify gait from non-gait. We observed that we missed more than half of the gait windows from the test set. We attribute this to the dominating presence of other acoustic events and background noise overlapping with footstep events in the melspectrograms which misled our judgments. Figure 3 shows four gait melspectrograms as an example in how noise and polyphony made it difficult for us to distinguish between the two types correctly. This led us to realise that we need a technique to remove the noisy melspectrograms from the training set for the model.

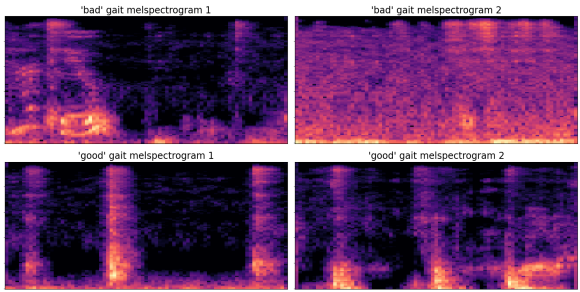

*Figure 3 Examples of suitable and unsuitable gait melspectrograms for training as observed in the data captured*

*D. Parameters for the filter*

To study the patterns between suitable and unsuitable gait melspectrograms, we manually went through the 1399 gait melspectrograms and annotated them between what we considered as clean and unclean/noisy melspectrograms as human experts. We labelled a total of 422 of them as "good" while the remaining 977 as "bad" gait melspectrograms. Next we devised the following set of 3 parameters guided by our observations during this manual labelling process. Figure 4 below visualises the parameters from two melspectrograms.

1. Average Peak Prominence: Gait melspectrograms suitable for training have distinct vertical peaks with little spectral noise around them. We capture this property using peak prominence in the energy signal representation of the melspectrogram. We define the energy signal from a melspectrogram as $E(i) = \sum_i F(i)$ for all frames in a melspectrogram where $F(i)$ is the $i$-th frame in the melspectrogram and $\sum_i F(i)$ is the sum of the amplitude magnitude values in each frame in decibels. Peak prominence measures how much a peak stands out in a signal due to its instristic height and its location relative to other peaks.

2. Residuals to the RMS: High amount of noise in a gait melspectrogram results in an energy signal $E$ with less prominent peaks for the individual foostep events. To capture this property, we computed the residuals of $E(i)$ for all frames i around the RMS of $E(i)$. We define this operation as $r = \sum_i (E(i) - RMS(E)^2)$ where $r$ is our second parameter.

3. Average Distance between Peaks: Footstep events appeared to occur with a rhythm within each gait melspectrogram. Non-gait melspectrograms which look similar to gait melspectrograms will not share this characteristic. We defined this parameter based on the average distance between all the peaks detected in the energy signal representation of the melspectrogram.

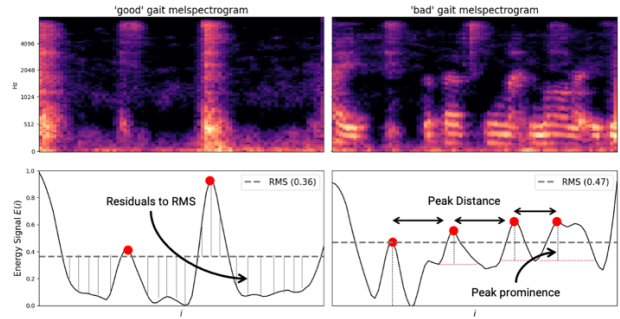

*Figure 4 Visual representation of the three parameters as extracted from the energy signal representation $E(i)$ of the melspectrogram*

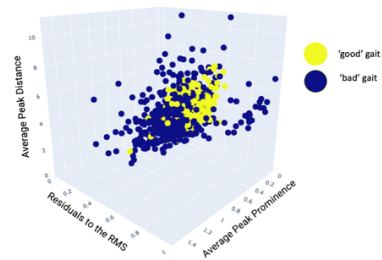

*Figure 5 The parameters combined in 3D space achieve separability for 'good' and 'bad' gait melspectrograms*

We extracted these 3 parameters across all of the 1399 spectrograms and visualised them in a third dimensional space to demonstrate that how it is able to successfully separate the suitable from the unsuitable gait melspectograms we found manually. Figure 5 visualises this "feature separability" which we sought to capture and automate from our manual effort.

*E. Training a machine learning algorithm as a filter*

We used Microsoft FLAML's AutoML [20] framework to automate finding the optimal ensemble of machine learning algorithms capable to performing the classification of the two types of gait melspectrograms we established earlier through manual annotation. Our focus was to maximise the predictive power of this model to detect the unsuitable melspectrograms as we have more of them in our training set to learn from. We took the original set of 1399 gait melspectrograms and our manual annotations and performed a 80-20 split again. The training set had 1119 melspectrograms with 338 labelled as "good" while the testing set had 280 melspectrograms with 84 labelled so. The AutoML model was trained on the 1119 melspectrograms with 10-fold cross-validation with the following parameters: `ensemble=True, metric="macro_f1", eval_method="cv", n_splits=10, time_budget=300, early_stop=True, estimator_list=["lgbm","xgboost","rf", "extra_tree", "kneighbour"], task = "classification"`. Table 1 below summarises the classification performance of the trained AutoML model as a filter on the test set of 280 melspectrogams against our manual annotations. Figure 6 shows the confusion matrix for then same.

TABLE I. Evaluating the AutoML filter against human expertise

| Gait melspec label | Precision | Recall | f(1) | samples |
|---|---|---|---|---|
| "bad" gait melspecs | 0.79 | 0.91 | 0.85 | 196 |
| "good" gait melspecs | 0.69 | 0.44 | 0.54 | 84 |
| **Average** | **0.76** | **0.77** | **0.75** | |

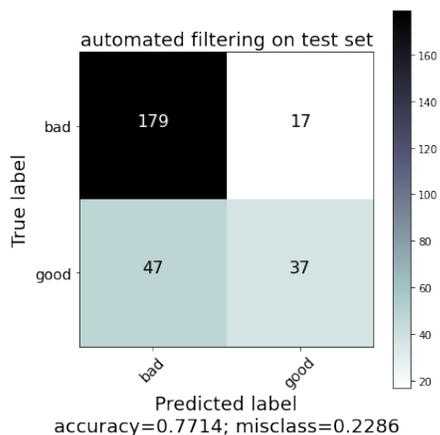

*Figure 6 Confusion Matrix for Table 1 above*

## IV. FILTER EFFECTIVENESS

We applied the newly trained AutoML model as a filter on the original "unfiltered" dataset of 3509 melspectrograms to get 2118 non-gait and 275 gait melspectrograms. To demonstrate the effectiveness of the filter as a way to remove unsuitable melspectrograms for gait detection models, we retrained the model from our previous work [6] on the filtered melspectrogram dataset and compared the performance on the test sets from the model trained in Section III(D) earlier. Figure 7 below visualises the improvements made by the new version of the model trained on the filtered gait melspectrograms across precision, recall and overall $f(1)$ scores.

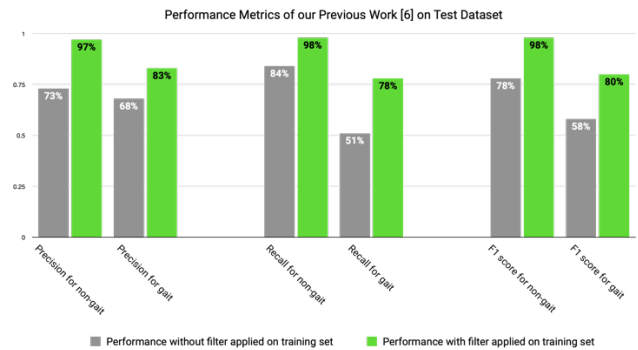

*Figure 7 Using the AutoML model as a filter on the training set improves performance metrics for gait and non-gait classification*

## V. CONCLUSION

This paper has presented an AutoML based classifier as an audio signal filter for gait data. We demonstrated how using the three parameters of average peak prominence, residuals around the RMS and the average peak distance together can help eliminate gait recordings which are noisy and are not suitable for training and developing robust acoustic based gait detection systems for older adults living in home and carehome settings. The proposed technique behind the filter when applied to the training set of a gait detector immediately improved its overall performance in terms of precision, recall and $f(1)$ score in gait detection.

We believe this filter will help in two areas of developing acoustic based gait detection systems for AGA in geriatrics: One, it will speed up checking human annotations for large-scale gait datasets and help researchers focus on good quality gait recordings; Second, it will help eliminate bad quality gait recordings directly when applied at the source of gait data collection.